\documentclass[aps,prc,twocolumn,preprintnumbers,superscriptaddress,nofootinbib]{revtex4-1}
\usepackage[utf8]{inputenc}
\usepackage{graphicx}
\usepackage{psfrag}
\usepackage{amssymb}
\usepackage{amsmath}
\usepackage{epstopdf}
\usepackage{color}
\usepackage{bm}% bold math
\usepackage{ulem}

\begin{document}

%\preprint{MSUCL xxx}

\title{Nuclear Fermi Momenta of $^{2}$H, $^{27}$Al and $^{56}$Fe from an Analysis of CLAS data}

\author{Hui Liu}
\email{liuhui@itp.ac.cn}
\affiliation{Institute of Modern Physics, Chinese Academy of Sciences, Lanzhou 730000, China}
\affiliation{School of Fundamental Physics and Mathematical Sciences, Hangzhou Institute for Advanced Study, UCAS, Hangzhou 310024, China}
%\affiliation{School of Nuclear Science and Technology, University of Chinese Academy of Sciences, Beijing 100049, China}

\author{Na-Na Ma}
\email{mann15@lzu.edu.cn}
\affiliation{School of Nuclear Science and Technology, Lanzhou University, Lanzhou 730000, China}

\author{Rong Wang}
\email{rwang@impcas.ac.cn (corresponding author)}
\affiliation{Institute of Modern Physics, Chinese Academy of Sciences, Lanzhou 730000, China}
\affiliation{School of Nuclear Science and Technology, University of Chinese Academy of Sciences, Beijing 100049, China}

\date{\today}

\begin{abstract}
Nuclear Fermi momentum is a basic property of a nucleus where many nucleons dwell.
However, in experiments only the nuclear Fermi momenta of just a few nuclei are measured
using quasielastic electron scattering on the nuclear targets so far.
Particularly, we still do not know experimentally the Fermi momentum
of the lightest nucleon composite -- the deuteron.
In this paper, we apply both gaussian distribution and Cauchy distribution
to describe the quasielastic peak in the cross section of electron-nucleus scattering.
The dip of the cross-section ratio at about $x_{\rm B}=1$ is
explained with the nuclear Fermi momentum.
By performing the least-square fits to the published CLAS data
in the narrow kinematic region of quasielastic scattering,
we obtain the nuclear Fermi momenta of $^{2}$H, $^{27}$Al and $^{56}$Fe,
which are $116\pm 7$ MeV/c, $232\pm 27$ MeV/c, and $244\pm 28$ MeV/c respectively.
The extracted nuclear Fermi momenta are compared to the simple calculations
based on Fermi gas model, and the consistencies are found.
\end{abstract}

%\pacs{25.40.Kk, 95.30.Cq}

\maketitle

\section{Introduction}
\label{sec:intro}

An atomic nucleus is a compact system composed of nucleon fermions
-- protons and neutrons -- under the strong nuclear force.
The Fermi statistic can be used to describe the single particle motion
inside the nucleus, and the nuclear Fermi momentum is used to describe
the independent nucleon energy level near Fermi surface.
In a nucleus, the nucleons move with an average momentum
which is closely related to the nuclear Fermi momentum.
The motions of the nucleons affect the scattering processes
under the high momentum transfer, such as the quasielastic scattering
\cite{Moniz:1969sr,Moniz:1971mt,Whitney:1974hr,Donnelly:1975ze,Benhar:2006wy},
the transverse momentum spectrum \cite{Fredriksson:1975tp,Fredriksson:1975cm,Yong:2015gma}
and the deep inelastic scattering \cite{Saito:1985ct,Arneodo:1992wf,Geesaman:1995yd,Malace:2014uea}.
What is more? The better understanding of the nuclear Fermi momentum
helps us better classifying the high-momentum tail of the nucleon
momentum distribution and the short-range correlations among the nucleons
\cite{Arrington:2011xs,Hen:2016kwk,Hen:2014nza,Schmidt:2020kcl,Wang:2020uhj},
since the quasielastic scattering is sensitive to both
the single nucleon motion and the nucleon-nucleon correlations.
For a better understanding of the high energy scattering experiments
on the nuclear targets, the nuclear Fermi motion effect is one of the corrections
that needs to be made.

The nuclear Fermi momentum can be deduced via the quasielastic scattering
of a high energy electron on the nuclear target \cite{Moniz:1969sr,Moniz:1971mt,Whitney:1974hr},
due to the weakly electromagnetic interaction which does not disturb much the structure of the target.
The quasielastic scattering can simply be viewed as the electron scattering
from an individual and moving nucleon in the Fermi sea,
with the recoiling nucleon going outside of the Fermi sphere
(escaping from the nucleus for most of the cases).
A quasielastic peak dominates in the spectrum of the electron energy loss
below the continuum region caused by deep inelastic scattering.
The width of the quasielastic peak is broadened by the Fermi motion
of the nucleon, and the position of the peak is influenced by
the separation energy between the struck nucleon and remanent nucleus \cite{Moniz:1969sr,Moniz:1971mt,Whitney:1974hr}.
Therefore measuring the width of the quasielastic peak is the key
of extracting the nuclear Fermi momentum.
Usually the differential cross section of quasielastic scattering
is described with the spectral function $S(\vec{k},E)$, where $\vec{k}$ and $E$
are the initial momentum and energy of the struck nucleon respectively \cite{Benhar:2006wy}.
In plane-wave-impulse approximation, the famous $y$-scaling
for quasielastic scattering implies the nucleonic degrees of freedom
and the initial nucleon momentum distribution \cite{Benhar:2006wy,West:1974ua,Sick:1980ey,Day:1987az,Arrington:1995hs}.

Extraction of the nuclear Fermi momentum gives us some guidance
in understanding the corresponding nuclear effect in high energy experiments.
To better understand the complex underlying neutrino-nucleus interaction,
physicists have measured the quasielastic scattering and resonance production
on the Titanium and Argon targets for the first time, at Hall A, JLab \cite{JeffersonLabHallA:2018zyx,Dai:2018gch}.
Recent high precision data of quasielastic scattering on the deuteron and some heavy nuclei
are provided by CLAS collaboration with the facilities at JLab \cite{Schmookler:2019nvf}.
The experiment was performed with 5 GeV electron beam hitting a dual target system,
with the scattered electrons measured with CLAS spectrometer.
Hence it is interesting to look at the CLAS data for quasielastic scattering peak
at $x_{\rm B}=\frac{Q^2}{2m\omega}\sim 1$,
where $Q^2$ and $\omega$ are the minus four-momentum square
and the energy of the exchanged virtual photon respectively.
The $e$-$A$ quasielastic cross sections are carefully given after
acceptance corrections, radiative corrections, Coulomb corrections,
and bin centering corrections. In the recent publication \cite{Schmookler:2019nvf},
the cross-section ratios are presented to quantify the nuclear effects,
which is also beneficial in reducing the systematic errors.

For the cross-section ratio of a heavy nucleus to deuteron,
it is interesting to note that there is a dip at $x_{\rm B}\sim 1$.
In this work, we demonstrate that the dip can be explained with the quasielastic peaks
of the heavy nucleus and the deuteron, which have quite different peak widths.
We use a simple Fermi smearing model in explaining the experimental data.
Both gaussian distribution and Cauchy distribution are applied
to describe the nuclear quasielastic scattering peak.
We extract the nuclear Fermi momenta of some unmeasured nuclei, especially the lightest
nucleon composite -- the deuteron.
In this work, the Fermi momentum of light nucleus is defined as the width of the peak
of nuclear cross section corresponding to quasi-elastic single-nucleon knock out.
This definition is regarded as an extension of the traditional picture
of Fermi momentum of heavy nucleus.
The method is discussed in Sec. \ref{sec:method}.
The determination of the deuteron Fermi momentum is presented in Sec. \ref{sec:deuteron}.
The determinations of the nuclear Fermi momenta
of aluminium and iron are presented in Sec. \ref{sec:Al-and-Fe}.
Some discussions and a concise summary is given in Sec. \ref{sec:summary}.

\section{Gaussian distribution and Cauchy distribution for nuclear quasielastic scattering peak}
\label{sec:method}

The quasielastic scattering on a nucleus is
viewed as the elastic scattering between the electron
probe and one nucleon inside the nucleus.
Compared to the $e$-$p$ elastic scattering, the energy transfer of quasielastic scattering
at a certain angle is smeared by the nucleon Fermi motion.
To depict the smearing from the randomly moving nucleon,
we apply the naive gaussian distribution and the long-tail Cauchy distribution for simplicity.
Cauchy distribution turns out to be a fine model description
in the narrow range around the quasielastic scattering peak.
As shown in the references \cite{Benhar:2015wva,Benhar:2015ula,Arrington:1998ps},
the inelastic scattering and nucleon-nucleon correlation
also contribute to the cross section around the quasi-elastic scattering peak,
but just a few percent in case of low-energy electron beams.

The differential cross section of nuclear quasielastic scattering over the energy transfer $\omega$
modeled with the gaussian distribution is given by,
\begin{equation}\label{eq:QECrosssection_gaussian}
\frac{1}{\mathrm{A}}\frac{d \sigma_\mathrm{A}(\omega)}{d\omega} =
\sigma_{\mathrm{eN}}\frac{1}{\sqrt{2 \pi}\zeta k_{\mathrm{F},\mathrm{A}}}
e^{-\frac{\left(\omega-\omega_{0}\right)^{2}}{2 \zeta^2 k_{\mathrm{F},\mathrm{A}}^{2}}},
\end{equation}
where $A$ is the mass number of the nucleus,
$\omega=E-E'$ is the energy transfer from the electron to the nucleon,
and $\sigma_{\mathrm{eN}}$ represents the total cross section of $e$-$N$ elastic scattering.
The main assumption of our model is that the width of the differential cross section peak
is proportional to the Fermi momentum $k_{\mathrm{F},\mathrm{A}}$
of the nucleon inside nucleus A \cite{Moniz:1969sr,Moniz:1971mt,Whitney:1974hr}.
Therefore, in Eq. (\ref{eq:QECrosssection_gaussian}), we introduce a pure coefficient $\zeta$,
which is a universal constant for all the nuclear targets in the same experiment.

In experiments, we usually present the cross section as a function of the Bjorken scaling variable $x_\mathrm{B}$.
In the nuclear target rest frame, the energy exchange $\omega$ is connected to $x_\mathrm{B}$,
via the definition of the Bjorken scaling variable.
$\omega$ is related to $x_\mathrm{B}$ as,
\begin{equation}\label{eq:omega_and-xB}
\omega=\frac{Q^{2}}{2 m x_{\mathrm{B}}},
\end{equation}
where $Q^2$ is the minus of the square of four-momentum transfer and $m$ is the nucleon mass.
Note that we see the quasielastic scattering peak of
the differential cross section with a slight shift away from $x_\mathrm{B}=1$,
and this shift of the peak comes from the average nucleon separation energy \cite{Moniz:1969sr,Moniz:1971mt,Whitney:1974hr}.
Again the gaussian smearing model does not work far from $x_\mathrm{B}=1$.
Using Eq. (\ref{eq:omega_and-xB}) and under the gaussian smearing model,
we can rewrite the per-nucleon quasielastic cross section
as a function of $x_\mathrm{B}$, which is written as,
\begin{equation}\label{eq:QECrosssection_dxB_gaussian}
\begin{split}
&\frac{1}{\mathrm{A}} \frac{d \sigma_\mathrm{A}(x_\mathrm{B})}{d x_\mathrm{B}} =
\left|\frac{1}{\mathrm{A}} \frac{d \sigma_\mathrm{A}(\omega)}{d \omega} \frac{d \omega}{d x_\mathrm{B}} \right| \\
&=\frac{\sigma_{\mathrm{e}\mathrm{N}}}{\sqrt{2 \pi}\zeta k_{\mathrm{F},\mathrm{A}}}
\exp\left[-\frac{(\frac{1}{x_\mathrm{B}}-X_0^\mathrm{A})^2}{2W_\mathrm{A}^2}\right]\frac{Q^2}{2 m x_\mathrm{B}^2}, \\
\end{split}
\end{equation}
where we have the parameter $X_0^\mathrm{A}=\frac{2 m \omega_{0}}{Q^2}$ for the central value
and the parameter $W_\mathrm{A}=\frac{2 m \zeta k_{\mathrm{F},\mathrm{A}}}{Q^2}$ for the width.
With $\mathrm{A}=2$ in Eq. (\ref{eq:QECrosssection_dxB_gaussian}), we get the quasielastic cross section for deuteron.
Dividing the quasielastic scattering cross section of a heavy nucleus by that of deuteron,
we then get a master formula for the cross-section ratio near $x_B\sim 1$,
\begin{equation}\label{eq:CrosssectionRatio_gaussian}
\left| \frac{d \sigma_\mathrm{A}/\mathrm{A}}{d \sigma_\mathrm{D}/2}\right| =
\frac{W_\mathrm{D}}{W_\mathrm{A}}\exp\left[\frac{(\frac{1}{x_\mathrm{B}}-X_0^\mathrm{D})^2}{2 W_\mathrm{D}^2}
-\frac{(\frac{1}{x_\mathrm{B}}-X_0^\mathrm{A})^2}{2 W_\mathrm{A}^2}\right].
\end{equation}
Here the subscript $\mathrm{D}$ denotes the deuterium.
Note that $W_\mathrm{A}=\frac{2 m \zeta k_{\mathrm{F},\mathrm{A}}}{Q^2}$ is linearly proportional
to the nuclear Fermi momentum of the nucleus A.
Once the width ratio $W_\mathrm{A_1}/W_\mathrm{A_2}$ is extracted from the data of the cross-section ratio,
the Fermi momentum ratio of the two nuclei is also obtained.
In Eq. (\ref{eq:CrosssectionRatio_gaussian}), the free parameter $X_0$ should be 1 for elastic scattering in theory.
Due to the mass deficit of the nucleon inside the nucleus,
the calculation of $x_{\mathrm{B}}$ should be different for the bound nucleon.
Hence the parameter $X_0$ is slightly away from 1.

In addition to the nuclear Fermi motion, the nucleon-nucleon short-range correlation
results in a small number of high momentum nucleons.
In order to model the long-tail momentum distribution from short-range correlations,
we apply a model using Cauchy distribution for the quasielastic scattering peak.
Similar to the derivations of the gaussian model, the quasielastic cross section
from Cauchy-distribution smearing is given by,
\begin{equation}\label{eq:QECrosssection_dxB_Cauchy}
\begin{split}
&\frac{1}{\mathrm{A}} \frac{d \sigma_\mathrm{A}(x_\mathrm{B})}{d x_\mathrm{B}} =
\left|\frac{1}{\mathrm{A}} \frac{d \sigma_\mathrm{A}(\omega)}{d \omega} \frac{d \omega}{d x_\mathrm{B}} \right| \\
&=\frac{\sigma_{\mathrm{e}\mathrm{N}}}{\pi\Gamma_\mathrm{A}}
\frac{1}{\left[ 1+\left(\frac{(1/x_\mathrm{B})-X_0^\mathrm{A}}{\Gamma_\mathrm{A}}\right)^2 \right]}  \frac{Q^2}{2 m x_\mathrm{B}^2}, \\
\end{split}
\end{equation}
in which the full width $\Gamma_{\rm A}$ is proportional to the Fermi momentum of nucleus $A$.
Then the cross-section ratio of a heavy nucleus $A$ to the deuteron ($D$) is written as,
\begin{equation}\label{eq:CrosssectionRatio_Cauchy}
\left| \frac{d \sigma_\mathrm{A}/\mathrm{A}}{d \sigma_\mathrm{D}/2}\right| =
\frac{\Gamma_\mathrm{D}}{\Gamma_\mathrm{A}}
\frac{\left[ 1+\left(\frac{(1/x_\mathrm{B})-X_0^\mathrm{D}}{\Gamma_\mathrm{D}}\right)^2 \right]}
{\left[ 1+\left(\frac{(1/x_\mathrm{B})-X_0^\mathrm{A}}{\Gamma_\mathrm{A}}\right)^2 \right]},
\end{equation}
which is the master equation in the Cauchy-distribution model.

The Fermi momenta of several nuclei were measured decades ago \cite{Moniz:1971mt}, such as $^{12}$C and $^{208}$Pb.
Recently the quasielastic cross-section ratios of some heavy nuclei ($^{12}$C, $^{27}$Al, $^{56}$Fe, and $^{208}$Pb)
to deuteron were analyzed by CLAS Collaboration in the data-mining project of 6 GeV data \cite{Schmookler:2019nvf}.
Therefore this provides us an opportunity to determine the Fermi momentum of deuteron,
with the models expressed in Eq. (\ref{eq:CrosssectionRatio_gaussian}) and Eq. (\ref{eq:CrosssectionRatio_Cauchy}),
combining the measurements of $^{12}$C and $^{208}$Pb.
The main objective of the paper is to provide the Fermi momenta of some unmeasured nuclei
based on the Fermi motion smearing model.

\section{Fermi momentum of deuteron}
\label{sec:deuteron}

\begin{figure}
	\begin{center}
		\includegraphics[width=0.85\linewidth]{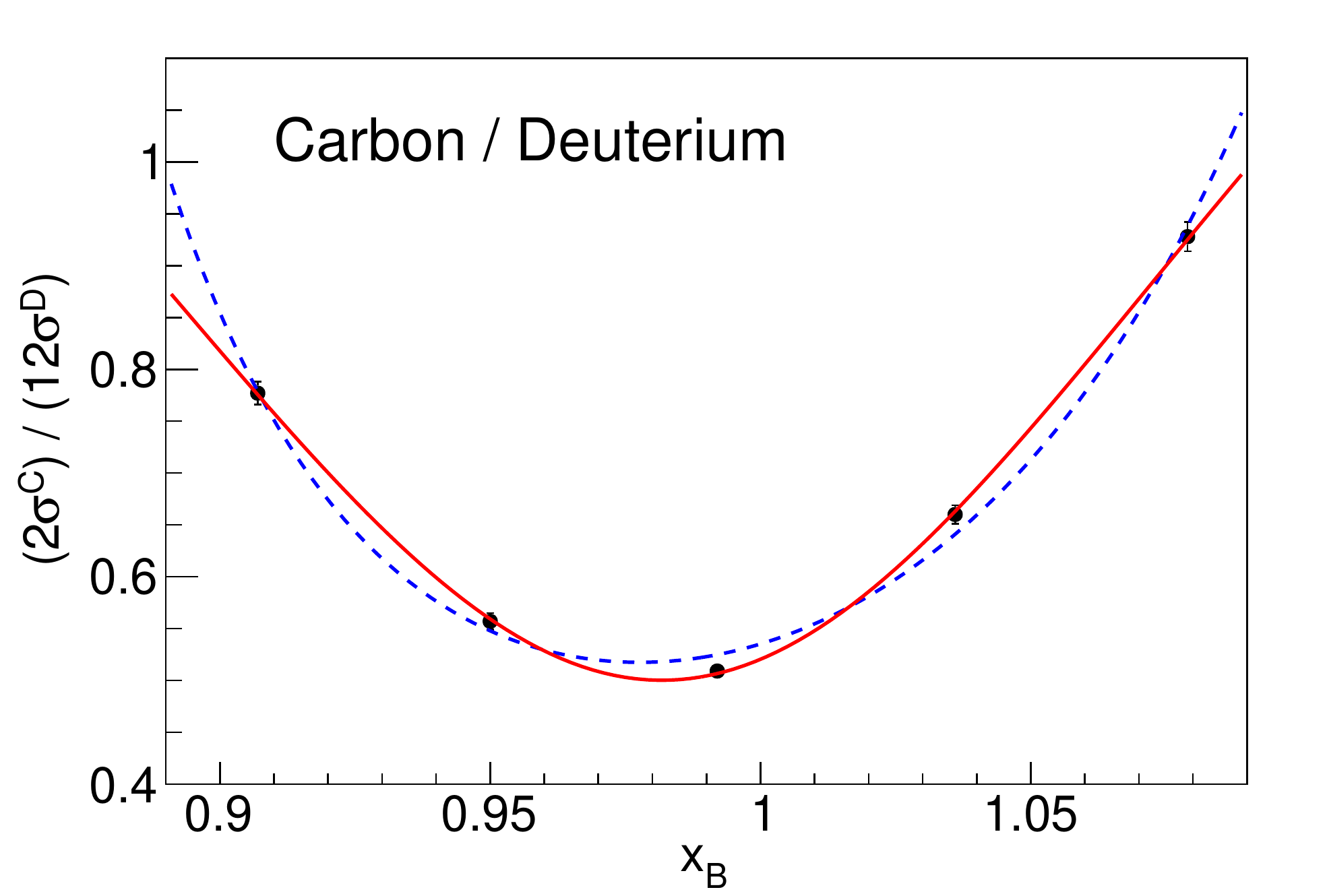}
		\caption{The quasielastic cross-section ratio $\left| \frac{d \sigma_\mathrm{A}/\mathrm{A}}{d \sigma_\mathrm{D}/2}\right|$ of carbon to deuteron
                 as a function of the Bjorken variable $x_\mathrm{B}$. The experimental data are taken from CLAS Collaboration ~\cite{Schmookler:2019nvf}.
                 The dashed blue curve shows a fit to a model of the assumption that the quasielastic peak is gaussian.
                 The solid red curve shows a fit to a model of the assumption that the quasielastic peak is Cauchy distribution.   }
		\label{fig:XsectionRatioCD}
	\end{center}
\end{figure}

\begin{figure}
	\begin{center}
		\includegraphics[width=0.85\linewidth]{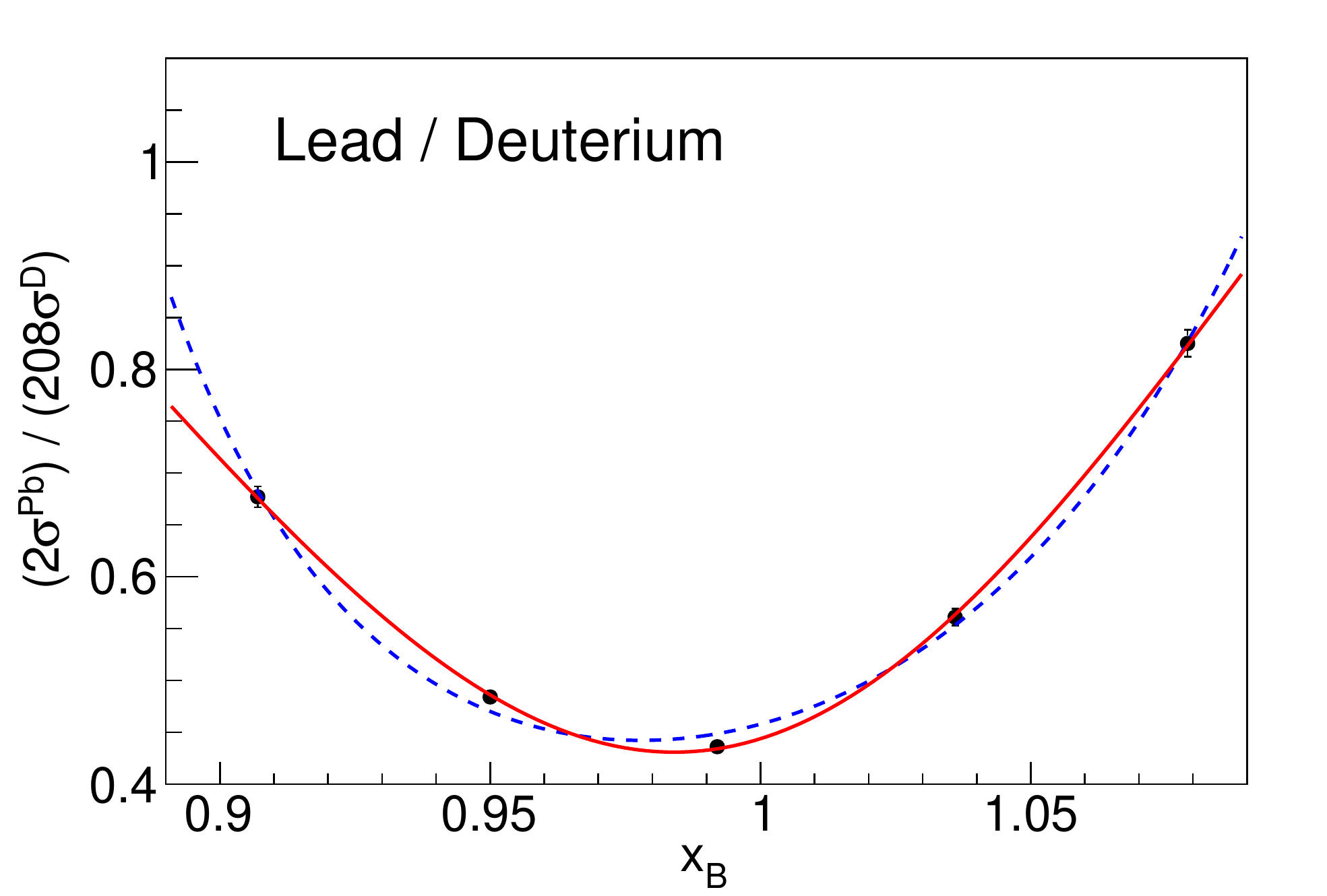}
		\caption{The quasielastic cross-section ratio $\left| \frac{d \sigma_\mathrm{A}/\mathrm{A}}{d \sigma_\mathrm{D}/2}\right|$ of lead to deuteron
                 as a function of the Bjorken variable $x_\mathrm{B}$. The experimental data are taken from CLAS Collaboration ~\cite{Schmookler:2019nvf}.
                 The dashed blue curve shows a fit to a model of the assumption that the quasielastic peak is gaussian.
                 The solid red curve shows a fit to a model of the assumption that the quasielastic peak is Cauchy distribution.   }
		\label{fig:XsectionRatioPbD}
	\end{center}
\end{figure}

Fig. \ref{fig:XsectionRatioCD} and Fig. \ref{fig:XsectionRatioPbD} show the per-nucleon
quasielastic cross-section ratios as a function of $x_\mathrm{B}$ \cite{Schmookler:2019nvf},
for $^{12}$C over deuteron
and for $^{208}$Pb over deuteron, respectively. The least-square fits based on the models
described in Eq. (\ref{eq:CrosssectionRatio_gaussian}) and Eq. (\ref{eq:CrosssectionRatio_Cauchy}) are performed to the data,
which are also shown in the figures.
The fitting results based on gaussian-distribution model and Cauchy-distribution model
are listed in Table \ref{tab:GaussModelFitResults} and Table \ref{tab:CauchyModelFitResults} respectively.
Judged by the $\chi^2/ndf$ value, the model based on Cauchy distribution is much better than
the model based on gaussian distribution.
This implies that there is the high-momentum tail for the nucleon momentum distribution,
which is unveiled directly in the short-range correlation experiments in the past two decades.
We find that the peak position is consistent with $X_0=1$ for the quasielastic scattering on the deuteron
while the peak position $X_0$ is slightly lower than 1 for the heavy nuclei.
This indicates that the separation energy of the nucleon inside the deuteron is negligible.
For the heavy nuclei, the nucleon separation energy is much larger.

The small reduced $\chi^2$ ($<1$) of the fits suggest that the quasielastic scattering model
based on Cauchy distribution is in good consistency with the experimental data (see Table \ref{tab:CauchyModelFitResults}).
Therefore we use this model to extract the nuclear Fermi momentum of the unmeasured deuteron.
In the fits, we let the full width $\Gamma_\mathrm{A}$, the full width ratio $\Gamma_\mathrm{D}/\Gamma_\mathrm{A}$,
the median $X_0^\mathrm{D}$ and the median $X_0^\mathrm{A}$ be the free parameters.
$\Gamma_\mathrm{D}/\Gamma_\mathrm{C}$ is determined to be 0.51 $\pm$ 0.03,
and $\Gamma_\mathrm{D}/\Gamma_\mathrm{Pb}$ is determined to be 0.46 $\pm$ 0.04.
For a fit of four free parameters, the $1\sigma$ uncertainties for
the determined parameters are defined by $\chi^2=\chi^2_{\rm min}+4.88$ \cite{James:1994vla}.

The ratio of the quasielastic peak width actually can be regarded as the ratio of the nuclear Fermi momentum.
As the nuclear Fermi momenta of carbon and lead
are already measured to be 221 $\pm$ 5 MeV/c and 265 $\pm$ 5 MeV/c respectively
from the previous experiments \cite{Moniz:1971mt},
we then deduce the nuclear Fermi momentum of deuteron to be 113 $\pm$ 8 MeV/c
and 122 $\pm$ 11 MeV/c from the above obtained ratios of quasielastic peak widths.
It is delighting to find that the values of the Fermi momentum of deuteron obtained from
the cross-section ratios of different nuclear targets
are consistent with each other within the uncertainty.
To get an average, the Fermi momentum of deuteron is
combined to be 116 $\pm$ 7 MeV/c.

Note that the designed resolution of CLAS drift chamber
for 1 GeV/c charged particle is $\delta p/p \leq 0.5\%$,
which is of quite high precision.
The measured elastic peak resolution is no greater than 16 MeV at beam energies of 2.4 and 4 GeV \cite{Mestayer:2000we}.
Therefore the detector resolution for $x_{\rm B}$ is about $\delta x_{\rm B} \sim (16{\rm MeV}) / (938{\rm MeV}) \sim 0.017$,
which is much smaller than the measured full width of nuclear quasielastic peak.
Moreover, the detector resolution is much smaller than the bin size of $x_{\rm B}$ as well.
Owing to the scattered electrons of high energy were measured precisely with
the drift chamber system of CLAS, we neglect the tiny detector resolution in the analysis.

In how large kinematic region that our models are valid and effective?
To answer this question, we have performed the fits to Carbon/Deutron data
in a wider kinematic range from $x_{\rm B}=0.84$ to $x_{\rm B}=1.14$.
The gaussian-distribution model gives the fitting quality $\chi^2/ndf=106/3=35$
and the Cauchy-distribution model gives the fitting quality $\chi^2/ndf=34/3=11$.
We see that our models do not fit well in a broader kinematic region of $x_{\rm B}$.
Nevertheless for the Cauchy-distribution model,
the determined free parameters in the wider fitting range are consistent
with the ones in the narrow fitting range we applied.
The obtained parameters are $\Gamma_{\rm C}=0.30\pm 0.01$, $\Gamma_{\rm D}/\Gamma_{\rm C}=0.49\pm 0.02$,
$X_0^{\rm D}=1.01\pm 0.01$ and $X_0^{\rm C}=0.99\pm 0.01$
from the fit in the wider kinematic region of $0.84<x_{\rm B}<1.14$.

\begin{table}[h]
	\caption{\label{tab:GaussModelFitResults}
             The least-square fitting results of the cross-section ratio data,
             within the model of gaussian distribution for the quasielastic scattering peak.
             The fit quality, the width, the width ratio and the central value of the peak are listed.   }
	\begin{tabular}{cccccc}
            \hline\hline
                Fit & $\frac{\chi^2}{ndf}$ & $W_\mathrm{A}$ & $W_\mathrm{D}/W_\mathrm{A}$ & $X_0^\mathrm{D}$ & $X_0^\mathrm{A}$     \\
			\hline
                C/D & $\frac{11}{1}$  & $0.14\pm0.03$ & $0.54\pm0.07$  &   $1.01\pm0.02$  &   $0.97\pm0.06$    \\
                Pb/D& $\frac{9.6}{1}$ & $0.14\pm0.23$ & $0.53\pm0.64$  &   $0.99\pm0.12$  &   $0.92\pm0.20$    \\
            \hline\hline
                Fit & $\frac{\chi^2}{ndf}$ & $W_\mathrm{D}$ & $W_\mathrm{A}/W_\mathrm{D}$ & $X_0^\mathrm{D}$ & $X_0^\mathrm{A}$     \\
            \hline
                Al/D& $\frac{7.0}{1}$  & $0.073\pm0.005$ & $1.9\pm0.3$  &   $1.01\pm0.02$  &   $0.97\pm0.06$    \\
                Fe/D& $\frac{6.6}{1}$  & $0.074\pm0.005$ & $2.0\pm0.3$  &   $1.01\pm0.02$  &   $0.96\pm0.06$    \\
            \hline\hline
	\end{tabular}
\end{table}

\begin{table}[h]
	\caption{\label{tab:CauchyModelFitResults}
             The least-square fitting results of the cross-section ratio data,
             within the model of Cauchy distribution for the quasielastic scattering peak.
             The fit quality, the full width, the full-width ratio and the central value of the peak are listed.   }
	\begin{tabular}{cccccc}
            \hline\hline
                Fit & $\frac{\chi^2}{ndf}$ & $\Gamma_\mathrm{A}$ & $\Gamma_\mathrm{D}/\Gamma_\mathrm{A}$ & $X_0^\mathrm{D}$ & $X_0^\mathrm{A}$     \\
			\hline
                C/D & $\frac{0.37}{1}$ & $0.31\pm0.02$ & $0.51\pm0.03$  &   $1.01\pm0.01$  &   $0.99\pm0.03$ \\
                Pb/D& $\frac{0.33}{1}$ & $0.37\pm0.03$ & $0.46\pm0.04$  &   $1.01\pm0.01$  &   $0.96\pm0.03$ \\
            \hline\hline
                Fit & $\frac{\chi^2}{ndf}$ & $\Gamma_\mathrm{D}$ & $\Gamma_\mathrm{A}/\Gamma_\mathrm{D}$ & $X_0^\mathrm{D}$ & $X_0^\mathrm{A}$     \\
            \hline
                Al/D& $\frac{1.7}{1}$  & $0.159\pm0.008$ & $2.0\pm0.2$  &   $1.01\pm0.01$  &   $0.98\pm0.03$    \\
                Fe/D& $\frac{1.2}{1}$  & $0.165\pm0.008$ & $2.1\pm0.2$  &   $1.01\pm0.01$  &   $0.98\pm0.03$    \\
            \hline\hline
	\end{tabular}
\end{table}

\begin{figure}
	\begin{center}
		\includegraphics[width=0.85\linewidth]{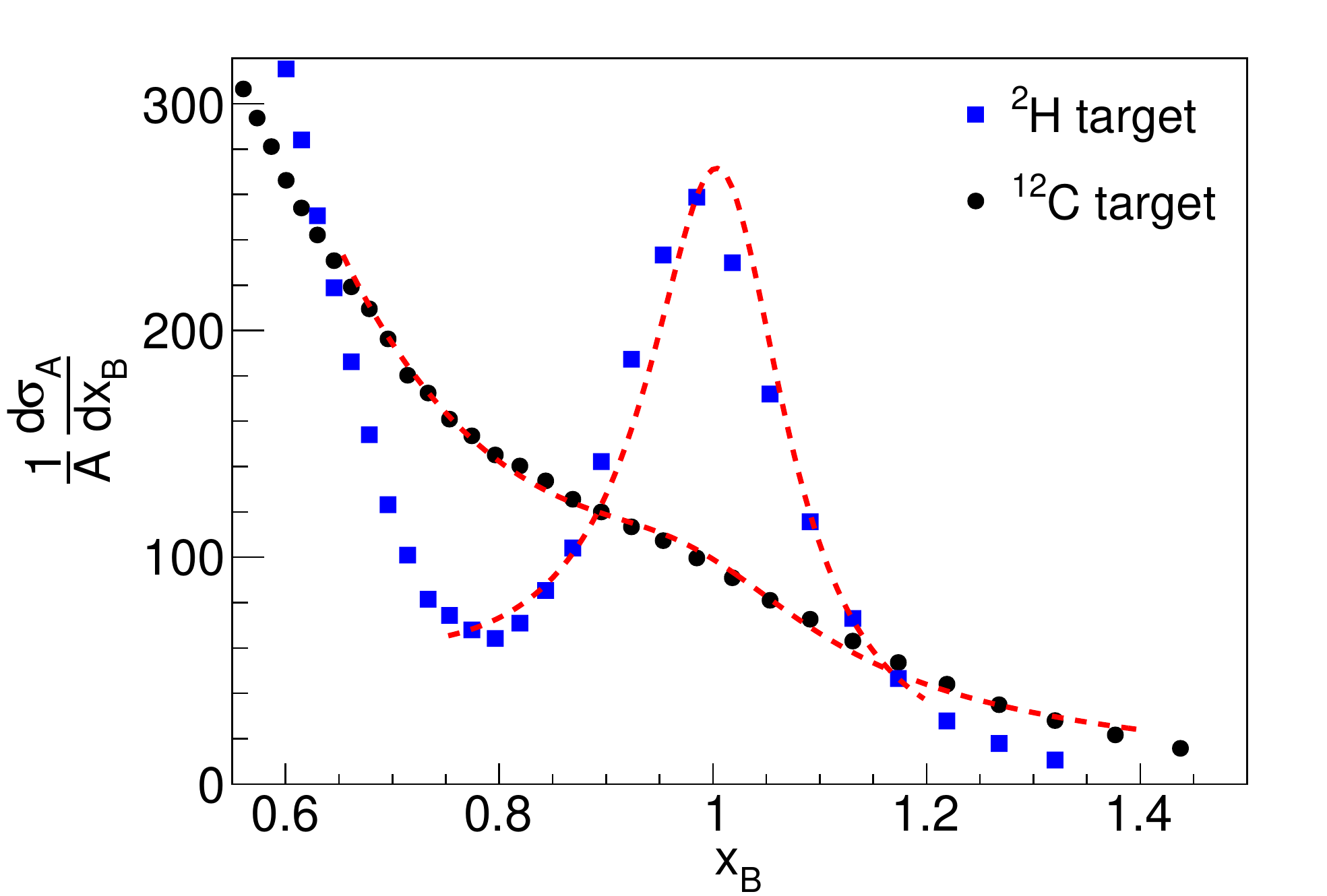}
		\caption{The electron inclusive cross sections on deuteron and carbon targets as a function of $x_{\rm B}$,
                 at the electron beam energy of 4.045 GeV.
                 The dashed curves show the fits based on the model of a Cauchy distribution
                 for quasi-elastic peak plus a continuum background for inelastic scattering.
                 For the fits, the full widths are fixed at 0.16 and 0.31 for deuteron and carbon respectively.
                 And the central values of the quasi-elastic peaks are fixed at 1.01 and 0.99 for deuteron and carbon respectively.   }
		\label{fig:NuclearXsections}
	\end{center}
\end{figure}

One of the main motivations for this study is to understand the dip presented
in the cross-section ratio around $x_{\rm B} = 1$.
In addition to the ratio between the cross sections of heavy nucleus and deuteron,
does the smearing model interpret the inclusive nuclear cross section as well?
Fig. \ref{fig:NuclearXsections} shows the electron inclusive scattering
cross sections of $^2$H and $^{12}$C as a function of $x_{\rm B}$.
The experimental data are taken from the references \cite{Arrington:1998ps,Benhar:2006er},
which were measured under the electron beam energy of 4.045 GeV.
Note that the $d\sigma/d\omega$ data were transformed into the $d\sigma/dx_{\rm B}$ data
according to Eq. (\ref{eq:QECrosssection_dxB_gaussian}).
The dashed curves show the fits of the Cauchy distribution for quasi-elastic
scattering peak plus the background for inelastic scattering and nucleon-nucleon correlations.
In performing the fits, the central values and the full widths of the Cauchy distributions
for deuteron and carbon are fixed as the fitted values listed in Table \ref{tab:CauchyModelFitResults}.
We find that the simple Cauchy-distribution model describes well the inclusive cross sections.
Unfixing the central values for Cauchy-distribution peaks,
the fits reproduce the experimental data even better.
We also find that there is significant contribution from inelastic scattering
for the carbon data at high energy around 4 GeV.
Therefore, to extract the nuclear Fermi momentum,
the low-energy data are preferred.

\section{Fermi momenta of aluminium and iron}
\label{sec:Al-and-Fe}

\begin{figure}
	\begin{center}
		\includegraphics[width=0.85\linewidth]{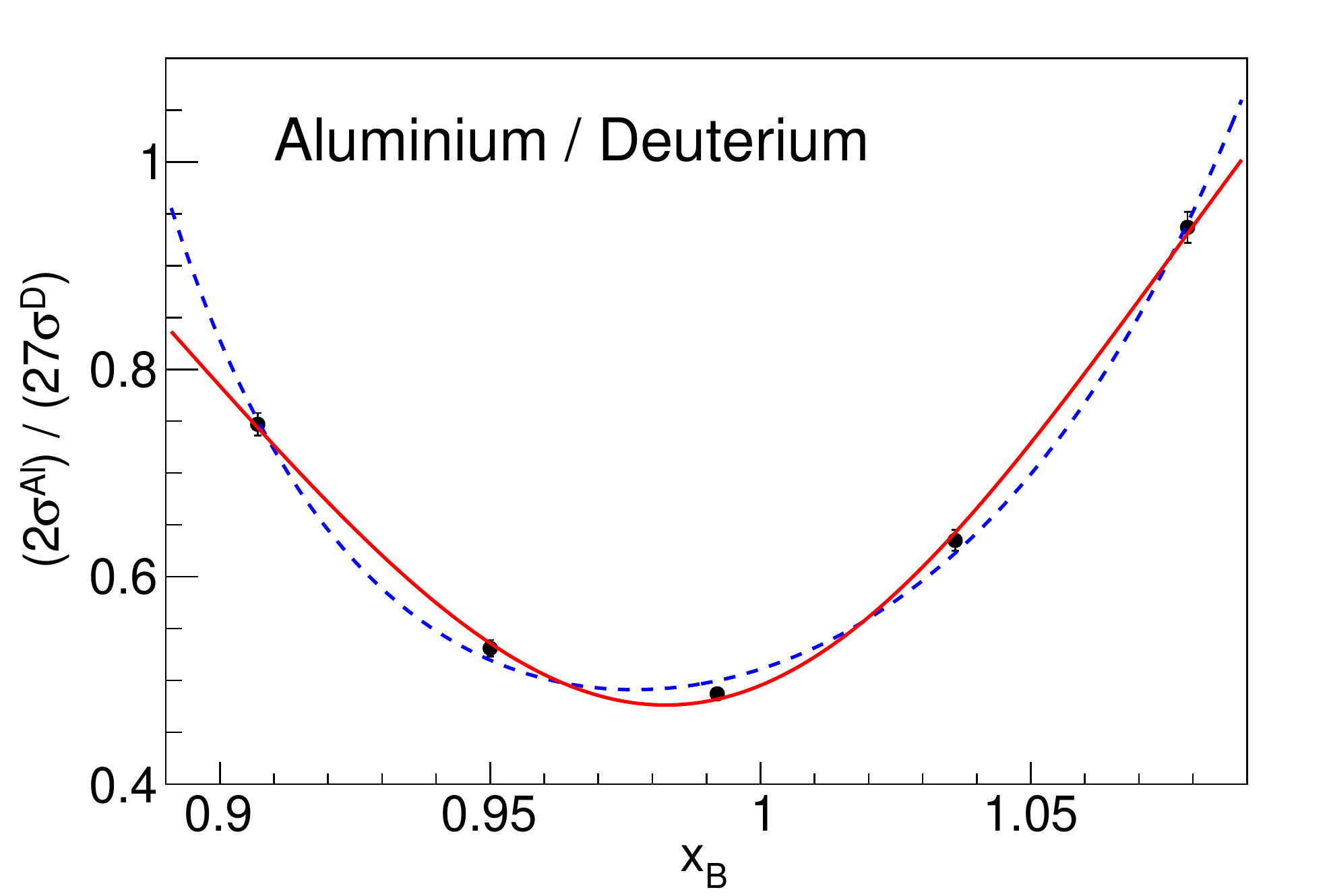}
		\caption{The quasielastic cross-section ratio $\left| \frac{d \sigma_\mathrm{A}/\mathrm{A}}{d \sigma_\mathrm{D}/2}\right|$ of aluminium to deuteron
                 as a function of the Bjorken variable $x_\mathrm{B}$. The experimental data are taken from CLAS Collaboration ~\cite{Schmookler:2019nvf}.
                 The dashed blue curve shows a fit to a model of the assumption that the quasielastic peak is gaussian.
                 The solid red curve shows a fit to a model of the assumption that the quasielastic peak is Cauchy distribution.   }
		\label{fig:XsectionRatioAlD}
	\end{center}
\end{figure}

\begin{figure}
	\begin{center}
		\includegraphics[width=0.85\linewidth]{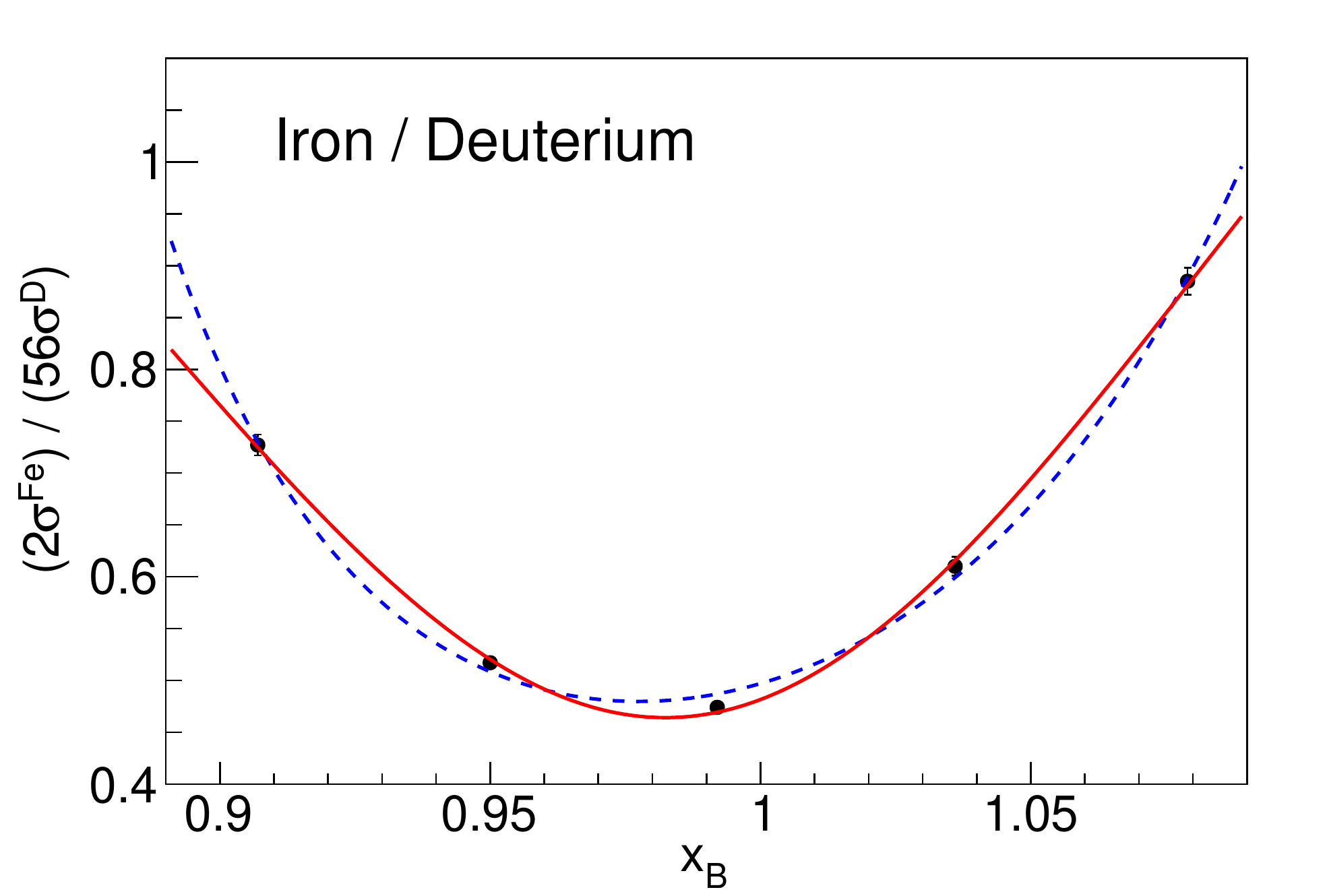}
		\caption{The quasielastic cross-section ratio $\left| \frac{d \sigma_\mathrm{A}/\mathrm{A}}{d \sigma_\mathrm{D}/2}\right|$ of iron to deuteron
                 as a function of the Bjorken variable $x_\mathrm{B}$. The experimental data are taken from CLAS Collaboration ~\cite{Schmookler:2019nvf}.
                 The dashed blue curve shows a fit to a model of the assumption that the quasielastic peak is gaussian.
                 The solid red curve shows a fit to a model of the assumption that the quasielastic peak is Cauchy distribution.   }
		\label{fig:XsectionRatioFeD}
	\end{center}
\end{figure}

The quasielastic scattering cross-section ratios between some other heavy nuclei
and the deuteron are also measured by CLAS Collaboration.
Fig. \ref{fig:XsectionRatioAlD} shows the cross-section ratio between $^{27}$Al and deuteron as a function of $x_{\rm B}$;
And Fig. \ref{fig:XsectionRatioFeD} shows the cross-section ratio between $^{56}$Fe and deuteron as a function of $x_{\rm B}$.
By performing the fits to the CLAS data within the model
of Cauchy-distribution peak (Eq. (\ref{eq:CrosssectionRatio_Cauchy})),
we have obtained the ratios of quasielastic peak widths:
$\Gamma_\mathrm{Al}/\Gamma_\mathrm{D}$ = 2.0 $\pm$ 0.2 and $\Gamma_\mathrm{Fe}/\Gamma_\mathrm{D}$ = 2.1 $\pm$ 0.2.
In Figs. \ref{fig:XsectionRatioAlD} and \ref{fig:XsectionRatioFeD},
we see that the Cauchy-distribution model describes the experimental data amazingly well,
with the $\chi^2/ndf$ around 1 (see Table \ref{tab:CauchyModelFitResults}).

The full-width ratio is directly equal to the nuclear-Fermi-momentum ratio.
As we have determined the Fermi momentum of the deuteron in the above section,
we can calculate the Fermi momentum of the heavy nucleus with
the obtained ratio of the Fermi momenta between the heavy nucleus and the deuteron.
Based on the results of the above fits with the Cauchy-distribution model,
we have determined the Fermi momenta
of $^{27}$Al and $^{56}$Fe, which are summarized in Table \ref{tab:FermiMomentumResults}
and comparable with the simple Fermi gas model calculations.

In a recent work \cite{Murphy:2019wed}, the Fermi momentum of $^{27}$Al
is estimated from the super scaling phenomenon around $\Psi\sim 0$.
Our obtained Fermi momentum of aluminum nucleus is obvious smaller than
the value given by the super scaling analysis \cite{Murphy:2019wed},
though the discrepancy between the two values is within the uncertainty of our analysis.
One possible source for the discrepancy is that the electron beam energies
and $Q^2$ are different for the data used in the two analyses,
which result in the different magnitudes of inelastic scattering contributions.
Ignoring the influence of inelastic scattering may introduce
a sizeable systematic uncertainties in our analysis.
The other possible source for the discrepancy is
the violation of super scaling phenomenon.

\begin{table}[h]
	\caption{\label{tab:FermiMomentumResults}
             Fermi Momenta of some nuclei determined in this work.
             $k_{\mathrm{F,~exp.}}$ denotes the Fermi momentum given by our analysis of the CLAS data.
             The errors are the statistical errors only.
             $k_{\mathrm{F,~theo.}}$ denotes the Fermi momentum given by the calculation
             from the Fermi gas model for the nucleus (see Eq. (\ref{eq:FerimiMomentum_density})).  }
	\begin{tabular}{ccccc}
            \hline\hline
			&Nucleus &$k_{\mathrm{F,~exp.}}$ (MeV/c)& $k_{\mathrm{F,theo.}}$ (MeV/c)&\\
			\hline
			&	$^{2}$H    &    $116\pm7$    &  140  &\\
			&	$^{27}$Al  &    $232\pm27$   &  226 &\\
			&	$^{56}$Fe  &    $244\pm28$   &  231 &\\
            \hline\hline
	\end{tabular}
\end{table}

\section{Discussions and summary}
\label{sec:summary}

From our analysis based on the simple model of nucleon momentum smearing, we find that
the nuclear Fermi momentum of deuteron is about half of that of a heavy nucleus,
while the Fermi momenta of aluminium and iron are close to the Fermi momentum of heavy nucleus
around 250 MeV/c. An interesting question is that whether or not
the Fermi motion of the nucleon in deuteron satisfies the Fermi gas model description.

In Fermi gas model, the Fermi momentum is directly connected to the nuclear density.
Based on Pauli exclusion principle,
the nucleon fermions fully occupy the quantum states in the nucleus.
Assuming a sphere nucleus, the number of nucleon fermions is simply counted as \cite{d2019introduction},
\begin{equation}\label{eq:nucleon_number}
\tilde{n} = 2 \frac{\left(\frac{4}{3}\pi R^3\right) \left(\frac{4}{3}\pi k_{\rm F}^3\right) }{(2\pi\hbar)^3 },
\end{equation}
where $\tilde{n}$ is the proton number or the neutron number,
$R$ is the radius of a nucleus and $k_{\rm F}$ is the Fermi momentum.
Approximately the nuclear radius is proportional to $A^{1/3}$.
With $R=r_0A^{1/3}$, the proton Fermi momentum $k_{\rm F}^{\rm p}$
in Fermi gas model is calculated as,
\begin{equation}\label{eq:FerimiMomentum_radius}
k_{\mathrm{F}}^{\mathrm{p}}=\frac{1}{r_0}\left(\frac{9}{4}\pi \frac{Z}{A}\right)^{\frac{1}{3}},
\end{equation}
in which $\hbar$ equals to 1 with the natural unit used.
Using the formula $\rho=A/(\frac{4}{3}\pi R^3)$, we can rewrite the proton Fermi momentum
as a function of the nuclear density $\rho$, which is written as,
\begin{equation}\label{eq:FerimiMomentum_density}
k_{\mathrm{F}}^{\mathrm{p}}=\left(3 \pi^2 \frac{Z}{A} \rho \right)^{\frac{1}{3}}.
\end{equation}
Here we assume a spherical nucleus and we set $\hbar$ to 1
with the natural unit used.

We adopt the nuclear density data from Ref. \cite{Gomez:1993ri} (see Table~\ref{tab:nuclear_density}
for the nuclei studied in the reference). These average nuclear densities are calculated
with $\rho(A) = 3A/(4\pi R_e^3)$ and $R_e^2 = 5\left<r^2\right>/3$.
$\sqrt{ \left<r^2\right> }$ is the root-mean-square radius
from the elastic electron scattering on the nucleus \cite{Gomez:1993ri}.

\begin{table}[h]
	\caption{\label{tab:nuclear_density} Nucleon densities of some nuclei \cite{Gomez:1993ri}. }
	\begin{tabular}{ccccccc}
            \hline\hline
			&A                        & $^{2}$H &  $^{12}$C & $^{27}$Al & $^{56}$Fe &\\
			\hline
			&$\rho$ (fm$^{-3}$)&    0.024  & 0.089      &    0.106     &    0.117      &\\
            \hline\hline
	\end{tabular}
\end{table}
With the nucleon density of $^{12}$C, the Fermi momentum of $^{12}$C is 216 MeV/c based on the Fermi gas model
described above, which is consistent with the previously measured value $221\pm 5$ MeV/c \cite{Moniz:1971mt}.
The Fermi gas model predictions for some other nuclei are also listed in Table \ref{tab:FermiMomentumResults}.
We find that the Fermi momenta of $^{27}$Al and $^{56}$Fe extracted in this work agree well with
the Fermi gas model calculations, while the Fermi momentum of deuteron is slightly smaller than the model prediction.
This implies that the Fermi gas model may not work well for the very light nucleus.
It is of no surprise because the deuteron is too light to fit in the Fermi gas picture.
Moreover, the determined Fermi momentum of deuteron is larger than the prediction 87 MeV/c
from a phenomenological parametrization \cite{Wang:2016mzo}, with
$k_{\mathrm{F}}(Z,N,A)=K_{\mathrm{F}}^{\mathrm{p}}(1-A^{-t_{\mathrm{p}}})\frac{Z}{A} + K_{\mathrm{F}}^{\mathrm{n}}(1-A^{-t_{\mathrm{n}}})\frac{N}{A}$.

The Fermi momentum of deuteron is extracted for the first time
from the quasielastic scattering data from CLAS \cite{Schmookler:2019nvf}.
The Fermi momentum of deuteron is $116\pm 7$ MeV/c, which is much smaller than that of the nucleus of high density.
The good fits of our model to the data indicate that the Cauchy distribution is valid
to fit the quasielastic cross section in a narrow range,
and the width of the quasielastic peak is proportional to the nuclear Fermi momentum.
The cross-section ratio for quasielastic scattering can be described with the formula in Eq. (\ref{eq:CrosssectionRatio_Cauchy}).
Within this formulism and the CLAS data, we have extracted the Fermi momenta of two
unmeasured heavy nuclei, $^{27}$Al and $^{56}$Fe.

The quasielastic scattering by the high energy electron beam on the nuclear target is
not only an important tool to see the short-range correlations among nucleons \cite{Schmookler:2019nvf},
but also a powerful method to acquire the information of nucleon Fermi motion inside the nucleus.
And these nucleon Fermi motion information would be very helpful
to the analyses in the experiments with the nuclear targets to study the properties of neutrons or protons
(such as the deuterium target).
Last but not least, to understand the Fermi momenta of the very light nuclei we need
more theoretical and experimental studies.

\begin{acknowledgments}
This work is supported by the National Natural Science Foundation of China under the Grant NO. 12005266
and the Strategic Priority Research Program of Chinese Academy of Sciences under the Grant NO. XDB34030301.
\end{acknowledgments}

\bibliographystyle{apsrev4-1}
\bibliography{TheReferences}

\begin{thebibliography}{34}%
\makeatletter
\providecommand \@ifxundefined [1]{%
 \@ifx{#1\undefined}
}%
\providecommand \@ifnum [1]{%
 \ifnum #1\expandafter \@firstoftwo
 \else \expandafter \@secondoftwo
 \fi
}%
\providecommand \@ifx [1]{%
 \ifx #1\expandafter \@firstoftwo
 \else \expandafter \@secondoftwo
 \fi
}%
\providecommand \natexlab [1]{#1}%
\providecommand \enquote  [1]{``#1''}%
\providecommand \bibnamefont  [1]{#1}%
\providecommand \bibfnamefont [1]{#1}%
\providecommand \citenamefont [1]{#1}%
\providecommand \href@noop [0]{\@secondoftwo}%
\providecommand \href [0]{\begingroup \@sanitize@url \@href}%
\providecommand \@href[1]{\@@startlink{#1}\@@href}%
\providecommand \@@href[1]{\endgroup#1\@@endlink}%
\providecommand \@sanitize@url [0]{\catcode `\\12\catcode `\$12\catcode
  `\&12\catcode `\#12\catcode `\^12\catcode `\_12\catcode `\%12\relax}%
\providecommand \@@startlink[1]{}%
\providecommand \@@endlink[0]{}%
\providecommand \url  [0]{\begingroup\@sanitize@url \@url }%
\providecommand \@url [1]{\endgroup\@href {#1}{\urlprefix }}%
\providecommand \urlprefix  [0]{URL }%
\providecommand \Eprint [0]{\href }%
\providecommand \doibase [0]{http://dx.doi.org/}%
\providecommand \selectlanguage [0]{\@gobble}%
\providecommand \bibinfo  [0]{\@secondoftwo}%
\providecommand \bibfield  [0]{\@secondoftwo}%
\providecommand \translation [1]{[#1]}%
\providecommand \BibitemOpen [0]{}%
\providecommand \bibitemStop [0]{}%
\providecommand \bibitemNoStop [0]{.\EOS\space}%
\providecommand \EOS [0]{\spacefactor3000\relax}%
\providecommand \BibitemShut  [1]{\csname bibitem#1\endcsname}%
\let\auto@bib@innerbib\@empty
%</preamble>
\bibitem [{\citenamefont {Moniz}(1969)}]{Moniz:1969sr}%
  \BibitemOpen
  \bibfield  {author} {\bibinfo {author} {\bibfnamefont {E.~J.}\ \bibnamefont
  {Moniz}},\ }\href {\doibase 10.1103/PhysRev.184.1154} {\bibfield  {journal}
  {\bibinfo  {journal} {Phys. Rev.}\ }\textbf {\bibinfo {volume} {184}},\
  \bibinfo {pages} {1154} (\bibinfo {year} {1969})}\BibitemShut {NoStop}%
%%CITATION = PHRVA,184,1154;%%
\bibitem [{\citenamefont {Moniz}\ \emph {et~al.}(1971)\citenamefont {Moniz},
  \citenamefont {Sick}, \citenamefont {Whitney}, \citenamefont {Ficenec},
  \citenamefont {Kephart},\ and\ \citenamefont {Trower}}]{Moniz:1971mt}%
  \BibitemOpen
  \bibfield  {author} {\bibinfo {author} {\bibfnamefont {E.~J.}\ \bibnamefont
  {Moniz}}, \bibinfo {author} {\bibfnamefont {I.}~\bibnamefont {Sick}},
  \bibinfo {author} {\bibfnamefont {R.~R.}\ \bibnamefont {Whitney}}, \bibinfo
  {author} {\bibfnamefont {J.~R.}\ \bibnamefont {Ficenec}}, \bibinfo {author}
  {\bibfnamefont {R.~D.}\ \bibnamefont {Kephart}}, \ and\ \bibinfo {author}
  {\bibfnamefont {W.~P.}\ \bibnamefont {Trower}},\ }\href {\doibase
  10.1103/PhysRevLett.26.445} {\bibfield  {journal} {\bibinfo  {journal} {Phys.
  Rev. Lett.}\ }\textbf {\bibinfo {volume} {26}},\ \bibinfo {pages} {445}
  (\bibinfo {year} {1971})}\BibitemShut {NoStop}%
%%CITATION = PRLTA,26,445;%%
\bibitem [{\citenamefont {Whitney}\ \emph {et~al.}(1974)\citenamefont
  {Whitney}, \citenamefont {Sick}, \citenamefont {Ficenec}, \citenamefont
  {Kephart},\ and\ \citenamefont {Trower}}]{Whitney:1974hr}%
  \BibitemOpen
  \bibfield  {author} {\bibinfo {author} {\bibfnamefont {R.~R.}\ \bibnamefont
  {Whitney}}, \bibinfo {author} {\bibfnamefont {I.}~\bibnamefont {Sick}},
  \bibinfo {author} {\bibfnamefont {J.~R.}\ \bibnamefont {Ficenec}}, \bibinfo
  {author} {\bibfnamefont {R.~D.}\ \bibnamefont {Kephart}}, \ and\ \bibinfo
  {author} {\bibfnamefont {W.~P.}\ \bibnamefont {Trower}},\ }\href {\doibase
  10.1103/PhysRevC.9.2230} {\bibfield  {journal} {\bibinfo  {journal} {Phys.
  Rev. C}\ }\textbf {\bibinfo {volume} {9}},\ \bibinfo {pages} {2230} (\bibinfo
  {year} {1974})}\BibitemShut {NoStop}%
\bibitem [{\citenamefont {Donnelly}\ and\ \citenamefont
  {Walecka}(1975)}]{Donnelly:1975ze}%
  \BibitemOpen
  \bibfield  {author} {\bibinfo {author} {\bibfnamefont {T.~W.}\ \bibnamefont
  {Donnelly}}\ and\ \bibinfo {author} {\bibfnamefont {J.~D.}\ \bibnamefont
  {Walecka}},\ }\href {\doibase 10.1146/annurev.ns.25.120175.001553} {\bibfield
   {journal} {\bibinfo  {journal} {Ann. Rev. Nucl. Part. Sci.}\ }\textbf
  {\bibinfo {volume} {25}},\ \bibinfo {pages} {329} (\bibinfo {year}
  {1975})}\BibitemShut {NoStop}%
\bibitem [{\citenamefont {Benhar}\ \emph {et~al.}(2008)\citenamefont {Benhar},
  \citenamefont {day},\ and\ \citenamefont {Sick}}]{Benhar:2006wy}%
  \BibitemOpen
  \bibfield  {author} {\bibinfo {author} {\bibfnamefont {O.}~\bibnamefont
  {Benhar}}, \bibinfo {author} {\bibfnamefont {D.}~\bibnamefont {day}}, \ and\
  \bibinfo {author} {\bibfnamefont {I.}~\bibnamefont {Sick}},\ }\href {\doibase
  10.1103/RevModPhys.80.189} {\bibfield  {journal} {\bibinfo  {journal} {Rev.
  Mod. Phys.}\ }\textbf {\bibinfo {volume} {80}},\ \bibinfo {pages} {189}
  (\bibinfo {year} {2008})},\ \Eprint {http://arxiv.org/abs/nucl-ex/0603029}
  {arXiv:nucl-ex/0603029} \BibitemShut {NoStop}%
\bibitem [{\citenamefont {Fredriksson}(1975)}]{Fredriksson:1975tp}%
  \BibitemOpen
  \bibfield  {author} {\bibinfo {author} {\bibfnamefont {S.}~\bibnamefont
  {Fredriksson}},\ }\href {\doibase 10.1016/0550-3213(75)90495-2} {\bibfield
  {journal} {\bibinfo  {journal} {Nucl. Phys. B}\ }\textbf {\bibinfo {volume}
  {94}},\ \bibinfo {pages} {337} (\bibinfo {year} {1975})}\BibitemShut
  {NoStop}%
\bibitem [{\citenamefont {Fredriksson}(1976)}]{Fredriksson:1975cm}%
  \BibitemOpen
  \bibfield  {author} {\bibinfo {author} {\bibfnamefont {S.}~\bibnamefont
  {Fredriksson}},\ }\href {\doibase 10.1016/0550-3213(76)90486-7} {\bibfield
  {journal} {\bibinfo  {journal} {Nucl. Phys. B}\ }\textbf {\bibinfo {volume}
  {111}},\ \bibinfo {pages} {167} (\bibinfo {year} {1976})}\BibitemShut
  {NoStop}%
\bibitem [{\citenamefont {Yong}(2017)}]{Yong:2015gma}%
  \BibitemOpen
  \bibfield  {author} {\bibinfo {author} {\bibfnamefont {G.-C.}\ \bibnamefont
  {Yong}},\ }\href {\doibase 10.1016/j.physletb.2016.12.013} {\bibfield
  {journal} {\bibinfo  {journal} {Phys. Lett. B}\ }\textbf {\bibinfo {volume}
  {765}},\ \bibinfo {pages} {104} (\bibinfo {year} {2017})},\ \Eprint
  {http://arxiv.org/abs/1503.08523} {arXiv:1503.08523 [nucl-th]} \BibitemShut
  {NoStop}%
\bibitem [{\citenamefont {Saito}\ and\ \citenamefont
  {Uchiyama}(1985)}]{Saito:1985ct}%
  \BibitemOpen
  \bibfield  {author} {\bibinfo {author} {\bibfnamefont {K.}~\bibnamefont
  {Saito}}\ and\ \bibinfo {author} {\bibfnamefont {T.}~\bibnamefont
  {Uchiyama}},\ }\href {\doibase 10.1007/BF01411895} {\bibfield  {journal}
  {\bibinfo  {journal} {Z. Phys. A}\ }\textbf {\bibinfo {volume} {322}},\
  \bibinfo {pages} {299} (\bibinfo {year} {1985})}\BibitemShut {NoStop}%
\bibitem [{\citenamefont {Arneodo}(1994)}]{Arneodo:1992wf}%
  \BibitemOpen
  \bibfield  {author} {\bibinfo {author} {\bibfnamefont {M.}~\bibnamefont
  {Arneodo}},\ }\href {\doibase 10.1016/0370-1573(94)90048-5} {\bibfield
  {journal} {\bibinfo  {journal} {Phys. Rept.}\ }\textbf {\bibinfo {volume}
  {240}},\ \bibinfo {pages} {301} (\bibinfo {year} {1994})}\BibitemShut
  {NoStop}%
\bibitem [{\citenamefont {Geesaman}\ \emph {et~al.}(1995)\citenamefont
  {Geesaman}, \citenamefont {Saito},\ and\ \citenamefont
  {Thomas}}]{Geesaman:1995yd}%
  \BibitemOpen
  \bibfield  {author} {\bibinfo {author} {\bibfnamefont {D.~F.}\ \bibnamefont
  {Geesaman}}, \bibinfo {author} {\bibfnamefont {K.}~\bibnamefont {Saito}}, \
  and\ \bibinfo {author} {\bibfnamefont {A.~W.}\ \bibnamefont {Thomas}},\
  }\href {\doibase 10.1146/annurev.ns.45.120195.002005} {\bibfield  {journal}
  {\bibinfo  {journal} {Ann. Rev. Nucl. Part. Sci.}\ }\textbf {\bibinfo
  {volume} {45}},\ \bibinfo {pages} {337} (\bibinfo {year} {1995})}\BibitemShut
  {NoStop}%
\bibitem [{\citenamefont {Malace}\ \emph {et~al.}(2014)\citenamefont {Malace},
  \citenamefont {Gaskell}, \citenamefont {Higinbotham},\ and\ \citenamefont
  {Cloet}}]{Malace:2014uea}%
  \BibitemOpen
  \bibfield  {author} {\bibinfo {author} {\bibfnamefont {S.}~\bibnamefont
  {Malace}}, \bibinfo {author} {\bibfnamefont {D.}~\bibnamefont {Gaskell}},
  \bibinfo {author} {\bibfnamefont {D.~W.}\ \bibnamefont {Higinbotham}}, \ and\
  \bibinfo {author} {\bibfnamefont {I.}~\bibnamefont {Cloet}},\ }\href
  {\doibase 10.1142/S0218301314300136} {\bibfield  {journal} {\bibinfo
  {journal} {Int. J. Mod. Phys. E}\ }\textbf {\bibinfo {volume} {23}},\
  \bibinfo {pages} {1430013} (\bibinfo {year} {2014})},\ \Eprint
  {http://arxiv.org/abs/1405.1270} {arXiv:1405.1270 [nucl-ex]} \BibitemShut
  {NoStop}%
\bibitem [{\citenamefont {Arrington}\ \emph {et~al.}(2012)\citenamefont
  {Arrington}, \citenamefont {Higinbotham}, \citenamefont {Rosner},\ and\
  \citenamefont {Sargsian}}]{Arrington:2011xs}%
  \BibitemOpen
  \bibfield  {author} {\bibinfo {author} {\bibfnamefont {J.}~\bibnamefont
  {Arrington}}, \bibinfo {author} {\bibfnamefont {D.~W.}\ \bibnamefont
  {Higinbotham}}, \bibinfo {author} {\bibfnamefont {G.}~\bibnamefont {Rosner}},
  \ and\ \bibinfo {author} {\bibfnamefont {M.}~\bibnamefont {Sargsian}},\
  }\href {\doibase 10.1016/j.ppnp.2012.04.002} {\bibfield  {journal} {\bibinfo
  {journal} {Prog. Part. Nucl. Phys.}\ }\textbf {\bibinfo {volume} {67}},\
  \bibinfo {pages} {898} (\bibinfo {year} {2012})},\ \Eprint
  {http://arxiv.org/abs/1104.1196} {arXiv:1104.1196 [nucl-ex]} \BibitemShut
  {NoStop}%
\bibitem [{\citenamefont {Hen}\ \emph {et~al.}(2017)\citenamefont {Hen},
  \citenamefont {Miller}, \citenamefont {Piasetzky},\ and\ \citenamefont
  {Weinstein}}]{Hen:2016kwk}%
  \BibitemOpen
  \bibfield  {author} {\bibinfo {author} {\bibfnamefont {O.}~\bibnamefont
  {Hen}}, \bibinfo {author} {\bibfnamefont {G.~A.}\ \bibnamefont {Miller}},
  \bibinfo {author} {\bibfnamefont {E.}~\bibnamefont {Piasetzky}}, \ and\
  \bibinfo {author} {\bibfnamefont {L.~B.}\ \bibnamefont {Weinstein}},\ }\href
  {\doibase 10.1103/RevModPhys.89.045002} {\bibfield  {journal} {\bibinfo
  {journal} {Rev. Mod. Phys.}\ }\textbf {\bibinfo {volume} {89}},\ \bibinfo
  {pages} {045002} (\bibinfo {year} {2017})},\ \Eprint
  {http://arxiv.org/abs/1611.09748} {arXiv:1611.09748 [nucl-ex]} \BibitemShut
  {NoStop}%
\bibitem [{\citenamefont {Hen}\ \emph {et~al.}(2014)\citenamefont {Hen} \emph
  {et~al.}}]{Hen:2014nza}%
  \BibitemOpen
  \bibfield  {author} {\bibinfo {author} {\bibfnamefont {O.}~\bibnamefont
  {Hen}} \emph {et~al.},\ }\href {\doibase 10.1126/science.1256785} {\bibfield
  {journal} {\bibinfo  {journal} {Science}\ }\textbf {\bibinfo {volume}
  {346}},\ \bibinfo {pages} {614} (\bibinfo {year} {2014})},\ \Eprint
  {http://arxiv.org/abs/1412.0138} {arXiv:1412.0138 [nucl-ex]} \BibitemShut
  {NoStop}%
\bibitem [{\citenamefont {Schmidt}\ \emph {et~al.}(2020)\citenamefont {Schmidt}
  \emph {et~al.}}]{Schmidt:2020kcl}%
  \BibitemOpen
  \bibfield  {author} {\bibinfo {author} {\bibfnamefont {A.}~\bibnamefont
  {Schmidt}} \emph {et~al.} (\bibinfo {collaboration} {CLAS}),\ }\href
  {\doibase 10.1038/s41586-020-2021-6} {\bibfield  {journal} {\bibinfo
  {journal} {Nature}\ }\textbf {\bibinfo {volume} {578}},\ \bibinfo {pages}
  {540} (\bibinfo {year} {2020})},\ \Eprint {http://arxiv.org/abs/2004.11221}
  {arXiv:2004.11221 [nucl-ex]} \BibitemShut {NoStop}%
\bibitem [{\citenamefont {Wang}\ \emph {et~al.}(2020)\citenamefont {Wang},
  \citenamefont {Thomas},\ and\ \citenamefont {Melnitchouk}}]{Wang:2020uhj}%
  \BibitemOpen
  \bibfield  {author} {\bibinfo {author} {\bibfnamefont {X.~G.}\ \bibnamefont
  {Wang}}, \bibinfo {author} {\bibfnamefont {A.~W.}\ \bibnamefont {Thomas}}, \
  and\ \bibinfo {author} {\bibfnamefont {W.}~\bibnamefont {Melnitchouk}},\
  }\href {\doibase 10.1103/PhysRevLett.125.262002} {\bibfield  {journal}
  {\bibinfo  {journal} {Phys. Rev. Lett.}\ }\textbf {\bibinfo {volume} {125}},\
  \bibinfo {pages} {262002} (\bibinfo {year} {2020})},\ \Eprint
  {http://arxiv.org/abs/2004.03789} {arXiv:2004.03789 [hep-ph]} \BibitemShut
  {NoStop}%
\bibitem [{\citenamefont {West}(1975)}]{West:1974ua}%
  \BibitemOpen
  \bibfield  {author} {\bibinfo {author} {\bibfnamefont {G.~B.}\ \bibnamefont
  {West}},\ }\href {\doibase 10.1016/0370-1573(75)90035-6} {\bibfield
  {journal} {\bibinfo  {journal} {Phys. Rept.}\ }\textbf {\bibinfo {volume}
  {18}},\ \bibinfo {pages} {263} (\bibinfo {year} {1975})}\BibitemShut
  {NoStop}%
\bibitem [{\citenamefont {Sick}\ \emph {et~al.}(1980)\citenamefont {Sick},
  \citenamefont {Day},\ and\ \citenamefont {Mccarthy}}]{Sick:1980ey}%
  \BibitemOpen
  \bibfield  {author} {\bibinfo {author} {\bibfnamefont {I.}~\bibnamefont
  {Sick}}, \bibinfo {author} {\bibfnamefont {D.}~\bibnamefont {Day}}, \ and\
  \bibinfo {author} {\bibfnamefont {J.~S.}\ \bibnamefont {Mccarthy}},\ }\href
  {\doibase 10.1103/PhysRevLett.45.871} {\bibfield  {journal} {\bibinfo
  {journal} {Phys. Rev. Lett.}\ }\textbf {\bibinfo {volume} {45}},\ \bibinfo
  {pages} {871} (\bibinfo {year} {1980})}\BibitemShut {NoStop}%
\bibitem [{\citenamefont {Day}\ \emph {et~al.}(1987)\citenamefont {Day} \emph
  {et~al.}}]{Day:1987az}%
  \BibitemOpen
  \bibfield  {author} {\bibinfo {author} {\bibfnamefont {D.~B.}\ \bibnamefont
  {Day}} \emph {et~al.},\ }\href {\doibase 10.1103/PhysRevLett.59.427}
  {\bibfield  {journal} {\bibinfo  {journal} {Phys. Rev. Lett.}\ }\textbf
  {\bibinfo {volume} {59}},\ \bibinfo {pages} {427} (\bibinfo {year}
  {1987})}\BibitemShut {NoStop}%
\bibitem [{\citenamefont {Arrington}\ \emph {et~al.}(1996)\citenamefont
  {Arrington} \emph {et~al.}}]{Arrington:1995hs}%
  \BibitemOpen
  \bibfield  {author} {\bibinfo {author} {\bibfnamefont {J.}~\bibnamefont
  {Arrington}} \emph {et~al.},\ }\href {\doibase 10.1103/PhysRevC.53.2248}
  {\bibfield  {journal} {\bibinfo  {journal} {Phys. Rev. C}\ }\textbf {\bibinfo
  {volume} {53}},\ \bibinfo {pages} {2248} (\bibinfo {year} {1996})},\ \Eprint
  {http://arxiv.org/abs/nucl-ex/9504003} {arXiv:nucl-ex/9504003} \BibitemShut
  {NoStop}%
\bibitem [{\citenamefont {Dai}\ \emph {et~al.}(2018)\citenamefont {Dai} \emph
  {et~al.}}]{JeffersonLabHallA:2018zyx}%
  \BibitemOpen
  \bibfield  {author} {\bibinfo {author} {\bibfnamefont {H.}~\bibnamefont
  {Dai}} \emph {et~al.} (\bibinfo {collaboration} {Jefferson Lab Hall A}),\
  }\href {\doibase 10.1103/PhysRevC.98.014617} {\bibfield  {journal} {\bibinfo
  {journal} {Phys. Rev. C}\ }\textbf {\bibinfo {volume} {98}},\ \bibinfo
  {pages} {014617} (\bibinfo {year} {2018})},\ \Eprint
  {http://arxiv.org/abs/1803.01910} {arXiv:1803.01910 [nucl-ex]} \BibitemShut
  {NoStop}%
\bibitem [{\citenamefont {Dai}\ \emph {et~al.}(2019)\citenamefont {Dai} \emph
  {et~al.}}]{Dai:2018gch}%
  \BibitemOpen
  \bibfield  {author} {\bibinfo {author} {\bibfnamefont {H.}~\bibnamefont
  {Dai}} \emph {et~al.},\ }\href {\doibase 10.1103/PhysRevC.99.054608}
  {\bibfield  {journal} {\bibinfo  {journal} {Phys. Rev. C}\ }\textbf {\bibinfo
  {volume} {99}},\ \bibinfo {pages} {054608} (\bibinfo {year} {2019})},\
  \Eprint {http://arxiv.org/abs/1810.10575} {arXiv:1810.10575 [nucl-ex]}
  \BibitemShut {NoStop}%
\bibitem [{\citenamefont {Schmookler}\ \emph {et~al.}(2019)\citenamefont
  {Schmookler} \emph {et~al.}}]{Schmookler:2019nvf}%
  \BibitemOpen
  \bibfield  {author} {\bibinfo {author} {\bibfnamefont {B.}~\bibnamefont
  {Schmookler}} \emph {et~al.} (\bibinfo {collaboration} {CLAS}),\ }\href
  {\doibase 10.1038/s41586-019-0925-9} {\bibfield  {journal} {\bibinfo
  {journal} {Nature}\ }\textbf {\bibinfo {volume} {566}},\ \bibinfo {pages}
  {354} (\bibinfo {year} {2019})}\BibitemShut {NoStop}%
%%CITATION = NATUA,566,354;%%
\bibitem [{\citenamefont {Benhar}\ \emph {et~al.}(2017)\citenamefont {Benhar},
  \citenamefont {Huber}, \citenamefont {Mariani},\ and\ \citenamefont
  {Meloni}}]{Benhar:2015wva}%
  \BibitemOpen
  \bibfield  {author} {\bibinfo {author} {\bibfnamefont {O.}~\bibnamefont
  {Benhar}}, \bibinfo {author} {\bibfnamefont {P.}~\bibnamefont {Huber}},
  \bibinfo {author} {\bibfnamefont {C.}~\bibnamefont {Mariani}}, \ and\
  \bibinfo {author} {\bibfnamefont {D.}~\bibnamefont {Meloni}},\ }\href
  {\doibase 10.1016/j.physrep.2017.07.004} {\bibfield  {journal} {\bibinfo
  {journal} {Phys. Rept.}\ }\textbf {\bibinfo {volume} {700}},\ \bibinfo
  {pages} {1} (\bibinfo {year} {2017})},\ \Eprint
  {http://arxiv.org/abs/1501.06448} {arXiv:1501.06448 [nucl-th]} \BibitemShut
  {NoStop}%
\bibitem [{\citenamefont {Benhar}\ \emph {et~al.}(2015)\citenamefont {Benhar},
  \citenamefont {Lovato},\ and\ \citenamefont {Rocco}}]{Benhar:2015ula}%
  \BibitemOpen
  \bibfield  {author} {\bibinfo {author} {\bibfnamefont {O.}~\bibnamefont
  {Benhar}}, \bibinfo {author} {\bibfnamefont {A.}~\bibnamefont {Lovato}}, \
  and\ \bibinfo {author} {\bibfnamefont {N.}~\bibnamefont {Rocco}},\ }\href
  {\doibase 10.1103/PhysRevC.92.024602} {\bibfield  {journal} {\bibinfo
  {journal} {Phys. Rev. C}\ }\textbf {\bibinfo {volume} {92}},\ \bibinfo
  {pages} {024602} (\bibinfo {year} {2015})},\ \Eprint
  {http://arxiv.org/abs/1502.00887} {arXiv:1502.00887 [nucl-th]} \BibitemShut
  {NoStop}%
\bibitem [{\citenamefont {Arrington}\ \emph {et~al.}(1999)\citenamefont
  {Arrington} \emph {et~al.}}]{Arrington:1998ps}%
  \BibitemOpen
  \bibfield  {author} {\bibinfo {author} {\bibfnamefont {J.}~\bibnamefont
  {Arrington}} \emph {et~al.},\ }\href {\doibase 10.1103/PhysRevLett.82.2056}
  {\bibfield  {journal} {\bibinfo  {journal} {Phys. Rev. Lett.}\ }\textbf
  {\bibinfo {volume} {82}},\ \bibinfo {pages} {2056} (\bibinfo {year}
  {1999})},\ \Eprint {http://arxiv.org/abs/nucl-ex/9811008}
  {arXiv:nucl-ex/9811008} \BibitemShut {NoStop}%
\bibitem [{\citenamefont {James}(1994)}]{James:1994vla}%
  \BibitemOpen
  \bibfield  {author} {\bibinfo {author} {\bibfnamefont {F.}~\bibnamefont
  {James}},\ }\href@noop {} {\  (\bibinfo {year} {1994})}\BibitemShut {NoStop}%
\bibitem [{\citenamefont {Mestayer}\ \emph {et~al.}(2000)\citenamefont
  {Mestayer} \emph {et~al.}}]{Mestayer:2000we}%
  \BibitemOpen
  \bibfield  {author} {\bibinfo {author} {\bibfnamefont {M.~D.}\ \bibnamefont
  {Mestayer}} \emph {et~al.},\ }\href {\doibase 10.1016/S0168-9002(00)00151-0}
  {\bibfield  {journal} {\bibinfo  {journal} {Nucl. Instrum. Meth. A}\ }\textbf
  {\bibinfo {volume} {449}},\ \bibinfo {pages} {81} (\bibinfo {year}
  {2000})}\BibitemShut {NoStop}%
\bibitem [{\citenamefont {Benhar}\ \emph {et~al.}(2006)\citenamefont {Benhar},
  \citenamefont {Day},\ and\ \citenamefont {Sick}}]{Benhar:2006er}%
  \BibitemOpen
  \bibfield  {author} {\bibinfo {author} {\bibfnamefont {O.}~\bibnamefont
  {Benhar}}, \bibinfo {author} {\bibfnamefont {D.}~\bibnamefont {Day}}, \ and\
  \bibinfo {author} {\bibfnamefont {I.}~\bibnamefont {Sick}},\ }\href@noop {}
  {\  (\bibinfo {year} {2006})},\ \Eprint
  {http://arxiv.org/abs/nucl-ex/0603032} {arXiv:nucl-ex/0603032} \BibitemShut
  {NoStop}%
\bibitem [{\citenamefont {Murphy}\ \emph {et~al.}(2019)\citenamefont {Murphy}
  \emph {et~al.}}]{Murphy:2019wed}%
  \BibitemOpen
  \bibfield  {author} {\bibinfo {author} {\bibfnamefont {M.}~\bibnamefont
  {Murphy}} \emph {et~al.},\ }\href {\doibase 10.1103/PhysRevC.100.054606}
  {\bibfield  {journal} {\bibinfo  {journal} {Phys. Rev. C}\ }\textbf {\bibinfo
  {volume} {100}},\ \bibinfo {pages} {054606} (\bibinfo {year} {2019})},\
  \Eprint {http://arxiv.org/abs/1908.01802} {arXiv:1908.01802 [hep-ex]}
  \BibitemShut {NoStop}%
\bibitem [{\citenamefont {D'Auria}(2019)}]{d2019introduction}%
  \BibitemOpen
  \bibfield  {author} {\bibinfo {author} {\bibfnamefont {S.}~\bibnamefont
  {D'Auria}},\ }\href@noop {} {\emph {\bibinfo {title} {Introduction to Nuclear
  and Particle Physics}}}\ (\bibinfo  {publisher} {Springer},\ \bibinfo {year}
  {2019})\BibitemShut {NoStop}%
\bibitem [{\citenamefont {Gomez}\ \emph {et~al.}(1994)\citenamefont {Gomez}
  \emph {et~al.}}]{Gomez:1993ri}%
  \BibitemOpen
  \bibfield  {author} {\bibinfo {author} {\bibfnamefont {J.}~\bibnamefont
  {Gomez}} \emph {et~al.},\ }\href {\doibase 10.1103/PhysRevD.49.4348}
  {\bibfield  {journal} {\bibinfo  {journal} {Phys. Rev. D}\ }\textbf {\bibinfo
  {volume} {49}},\ \bibinfo {pages} {4348} (\bibinfo {year}
  {1994})}\BibitemShut {NoStop}%
\bibitem [{\citenamefont {Wang}\ \emph {et~al.}(2017)\citenamefont {Wang},
  \citenamefont {Chen},\ and\ \citenamefont {Fu}}]{Wang:2016mzo}%
  \BibitemOpen
  \bibfield  {author} {\bibinfo {author} {\bibfnamefont {R.}~\bibnamefont
  {Wang}}, \bibinfo {author} {\bibfnamefont {X.}~\bibnamefont {Chen}}, \ and\
  \bibinfo {author} {\bibfnamefont {Q.}~\bibnamefont {Fu}},\ }\href {\doibase
  10.1016/j.nuclphysb.2017.04.008} {\bibfield  {journal} {\bibinfo  {journal}
  {Nucl. Phys. B}\ }\textbf {\bibinfo {volume} {920}},\ \bibinfo {pages} {1}
  (\bibinfo {year} {2017})},\ \Eprint {http://arxiv.org/abs/1611.03670}
  {arXiv:1611.03670 [hep-ph]} \BibitemShut {NoStop}%
\end{thebibliography}%
\end{document}